\documentclass[english,letterpaper,showpacs,floatfix]{revtex4}
\usepackage{epsfig}
\begin{document}
\newcommand{\be}{\begin{equation}}
\newcommand{\ee}{\end{equation}}
\newcommand{\bear}{\begin{eqnarray}}
\newcommand{\eear}{\end{eqnarray}}
\def\bearst{\begin{eqnarray*}}
\def\eearst{\end{eqnarray*}}
\title{ Brane World Cosmological Perturbations }

\author{Adenauer G. Casali}
\email{casali@fma.if.usp.br}
\author{ Elcio Abdalla}
\email{ eabdalla@fma.if.usp.br}
\affiliation{Instituto De Fisica, Universidade De Sao Paulo, C.P.66.318, CEP
05315-970,  Sao Paulo, Brazil }
\author{Bin Wang}
\email{binwang@fudan.ac.cn}
\affiliation{ Department of Physics, Fudan University, Shanghai 200433,
P. R. China }


\begin{abstract}
We consider a brane world and its gravitational linear perturbations. We
present a general solution of the perturbations in the bulk and find the
complete perturbed junction conditions for generic brane dynamics. We also
prove that (spin 2) gravitational waves in the great majority of cases can only
arise in connection with a non-vanishing anisotropic stress. This has far
reaching consequences for inflation in the brane world. Moreover, contrary
to the case of the radion, perturbations are stable.
\end{abstract}
\maketitle

\section{Introduction}
The brane world scenario is a new way of looking at the Universe from
the point of view of string/membrane theory, which has the advantage of
offering the opportunity of comparing prediction to observational results,
an aim that was beyond the scope of this area of theoretical physics for
more than two decades \cite{braneworld,venezia,polchinski}.

String theory is naturally defined in a higher dimensional space-time. 
It has been shown that string theory in various dimensions display very
natural behaviour under duality, which leads to the result that there is a
theory in 11 dimensions from which all string theories derive in a natural
way \cite{wittenstvariousd}. 

It is thus by now a widespread idea, from a general theoretical setup,
that the so-called M-theory \cite{polchinski} is a reasonable description
of our Universe: in the field theory limit, it is described by a solution
of the (eventually 11-dimensional) Einstein equations with a cosmological
constant, by means of a four dimensional membrane. In this picture only
gravity survives in the higher dimensions, while the remaining matter and
gauge interactions are typically four dimensional. This realizes the idea
that the real world is a membrane imbeded in a higher dimensional
space-time, such that the Standard model fields live in the membrane,
while gravity can enter into the bulk of the extra dimensions
\cite{horavawitten}. 

Randall and Sundrum \cite{randallsundrum} proposed a model of that type,
namely there exists a Universe described inside a membrane immersed in a
higher dimensional space-time, such that a so-called warp in the extra
dimensions prevents information to leak at large quantities outside of the
real world. 

A very large amount of new physics emerges. In particular, when membranes are
solutions of Einstein's equations and matter fields reside inside the brane,
the gravitational fields have to obey the Israel conditions \cite{israel}
at the sides of the brane. Thus, there is a possibility that
gravitational fields propagating out of the brane speed up, reaching
farther distances as compared to light propagating inside the brane, a
scenario that for a resident of the brane (such as ourselves) implies
shortcuts \cite{Abdalla2,csaki3,Ishihara,Caldwell,Chung}. Moreover, the
fact that the world is higher dimensional, together with general
properties of quantum field theory and further input such as the
holographic principle, indicate that further conclusions can be
drawn \cite{radion,Abdalla1,generalbranes,transdimen}.

In particular, due to the presence of the extra dimension, it is very
difficult to make predictions about the cosmological consequences by
studying the cosmological perturbations in the brane world. As the
simplest case, the cosmological perturbations were investigated by
neglecting the non-trivial evolution of perturbations in the bulk
\cite{maartensetal}. In order to solve the perturbation
equations including the bulk, a simplified inflation model,
in which de Sitter stage of inflation is instantaneusly connected to
Minskowski space, was considered in the study of gravitational wave
perturbations \cite{jhep2001}. Further applications of the approach of
employing conformally minkowskian coordinates can be found in
\cite{kodamaetal,Deruelle}. Using the conformally minkowskian coordinates,
Derulle et al \cite{Deruelle} disentangled the contributions of 
the bulk gravitons and of the motion of the brane, found the restrictions
of the bulk gravitons when matter on the brane is taken to be scalar and
solved analytically the brane perturbation equations. In this paper we are
going to generalize Deruelle et al's work with a full treatment of the
junction conditions leading to  the equations of motion on the brane, and
making direct connection with the standard perturbation theory.

Our aim here is to further develop these ideas by the description of
gravitational waves generated in early stages of the Universe. These are
important tools for the understanding of the model, since they can
actually be observed in the near future. They can also leave their
imprints in the background radiation and possibly be described by the
results of the Planck satelite.

\section{The scenario}

We shall thus consider gravitational perturbations of a five dimensional bulk where our Universe is described
by a four dimensional membrane with matter fields.
We consider a scenario where the unperturbed bulk is a purely Anti-de-Sitter 
space-time described by a metric conformally equivalent to minkowskian 
coordinates, that is, \cite{Deruelle},
\be
ds^{2} = g_{AB}dx^{A}dx^{B} = \frac{l^2}{(X^{4})^{2}}\eta_{AB}dX^{A}dX^{B},
\ee
with $l$ a length scale later called the Randall-Sundrum
scale and $\eta_{AB}$ the Minkowski metric. The notation is described in the appendix A.

We define the brane by a hyper-surface moving in the bulk with
$X^{0} = \bar{X}^{0}(\eta) = T(\eta)$, $X^{i} = \bar{X}^{i} = x^{i}$,
$X^{4} = \bar{X}^{4}(\eta) = A(\eta)$,
where $\eta$ is defined by $T'(\eta) = \sqrt{1+(A'(\eta))^{2}}$.

Parametrized in this way, the induced metric on the brane reads
\be
ds_{b}^{2} = \Bigl(\frac{l}{A(\eta)}\Bigr)^{2}(-d\eta^{2} +
\delta_{ij}dx^{i}dx^{j}) \quad .
\ee
Thus, $\eta$ is the conformal time of a FRW brane with
the scale-factor given by  $a(\eta) = l/A(\eta)$.

The components of the tangent vectors are
$\bar{V}_{0}^{A} = \frac{\partial \bar{X}^{A}}{\partial \eta} = (T'(\eta),
0, 0, 0, A'(\eta))$ and
$\bar{V}_{i}^{A} = \frac{\partial \bar{X}^{A}}{\partial x^{i}} = (0,
\delta^{A}_{i},0 )$, while the normal to the brane is $
\bar{n}^{A} = \Bigl(\frac{AA'}{l},0,0,0,\frac{AT'}{l}\Bigr) $. It obeys
the normalization condition $g_{AB}|_{\bar{b}}\bar{n}^{A}\bar{n}^{B} = 1$.

The energy-momentum tensor $\Pi_{AB}$ in the bulk is just a negative
cosmological constant . Therefore $\Pi_{AB}V^{A}_{\mu}n^{B} =0$  on the
brane. It can thus be shown that the junction conditions for a $Z_{2}$
symmetric brane imply the conservation of the brane energy-momentum, 
\be
\bar{\cal{T}}^{\mu}_{\nu ;\mu} = 0\quad .
\ee
On the other hand, the spatial part of the junction conditions
\cite{israel} reads 
\be
\frac{\kappa}{2}\Bigl(\bar{\cal{T}}^{i}_{j} -\frac{1}{3}
\delta^{i}_{j}\bar{\cal{T}}\Bigr) =\bar{K}^{i}_{j}\quad , \label{unKij}
\ee
where $\bar{\cal{T}}^{\mu}_{\nu}$ is the brane energy-momentum tensor and
$\bar{K}^{\mu}_{\nu}$ is the second fundamental form calculated on the
brane,
\be
\bar{K}_{\mu\nu} = -
\bar{V}^{A}_{\mu}\bar{V}^{B}_{\nu}\nabla_{A}\bar{n}_{B}|_{\bar{b}}\quad .  
\label{Kmunu}
\ee

For reference, we list the non-zero results for the background second
fundamental form on the brane,
\bear
K_{\eta\eta} &=&  \frac{l}{A^{2}T'}\Bigl(AA'' - T'^{2} \Bigr)\quad , \\ 
K_{ij} &=& \frac{lT'\delta_{ij}}{A^{2}}\quad .
\label{Kcompts}
\eear

For an isotropic and homogeneous distribution of matter in the brane
with a tension $\sigma$, we get the usual energy conservation equation
$\frac{d\rho}{d \tau} + 3H(\rho+P) = 0$,
where $d\tau = a(\eta) d\eta$ and $H = \dot{a}/a = \frac{1}{a}\frac{d
a}{d\tau} = a'(\eta)/a^2 = -A'(\eta)/l$, and, from the spatial part we
obtain the modified Friedmann equation
\be
H^{2} = \kappa^{2}(\rho + \sigma)^{2} - \frac{1}{l^{2}} =
\Bigl(\kappa^{2}\sigma^{2}-\frac{1}{l^2} \Bigr) + 6\kappa^{2}\sigma
\Bigl(\frac{\rho}{3}+\frac{\rho^{2}}{6\sigma} \Bigr)\quad .
\ee

This sets the whole unperturbed scenario. The next step is to consider the
perturbations and the subsequent wave equations.

\section{Perturbation}
Since perturbations are being treated linearly, it is possible to separate
the bulk perturbation from the perturbation of the position of the brane
and of the junction conditions, what we prefer to do for the sake of simplicity.

\subsection{Bulk Perturbation}
We first perturb the AdS bulk space-time. For general perturbations in 
conformally minkowskian coordinates \cite{Deruelle} the metric reads
\be
ds^{2} = \Bigl(\frac{l}{X^{4}}\Bigr)^{2}\Bigl(\eta_{AB} +
h_{AB}\Bigr)dX^{A}dX^{B}\quad ,
\ee
while on the brane we have
\be
ds_{b}^{2} = \frac{l^{2}}{A^2}(\eta_{\mu\nu} +
\gamma^{(b)}_{\mu\nu})dx^{\mu}dx^{\nu}\quad , 
\ee
where 
\bearst 
\gamma^{(b)}_{\eta\eta} &=& T'^{2}h_{00}|_{\bar{b}} +
A'^2h_{44}|_{\bar{b}} + 2T'A'h_{04}|_{\bar{b}}\quad ,\\ 
\gamma^{(b)}_{\eta i} &=& T'h_{0i}|_{\bar{b}} + A'h_{4i}|_{\bar{b}}\quad ,\\
\gamma^{(b)}_{ij} &=& h_{ij}|_{\bar{b}}\quad .
\eearst

Since we leave the perturbation in the position of the brane to a later 
stage, the tangent vectors to the brane are still the same as before, 
but we must find the correction to the normal to the brane. Writing the
perturbed normal as an unperturbed part calculated on the brane,
$\bar{n}^{A}$, and a small perturbation, $\delta n^{A}$, we have
\bearst
\eta_{AB}\bar{n}^{B}\delta n^{A} + \eta_{AB}\bar{n}^{A}\delta n^{B} +
h_{AB}|_{\bar{b}}\bar{n}^{B} \bar{n}^{A} &=& 0\quad ,\\ 
\eta_{AB}\bar{V}^{B}_{\mu}\delta n^{A} +  h_{AB}|_{\bar{b}}\bar{n}^{A}
\bar{V}^{B}_{\mu} &=& 0\quad .
\eearst

After some computation, we find for the covariant perturbation of the normal,
\be
\delta n_{A} = \frac{l^{2}}{A^{2}}h_{AB}|_{\bar{b}}\bar{n}^{B} +
\frac{l^{2}}{A^{2}}\eta_{AB}\delta n^{B}\quad ,
\ee
or more explicitly,
\bearst
\delta n_{0} &=& -h_{00}|_{\bar{b}}\frac{lA'^3}{2A} +
h_{04}|_{\bar{b}}\frac{lT'}{A}(1-A'^{2}) -
h_{44}|_{\bar{b}}\frac{T'^{2}A'l}{2A}, \\ 
\delta n_{4} &=& h_{00}|_{\bar{b}}\frac{lA'^{2}T'}{2A} +
h_{04}|_{\bar{b}}\frac{A'^{3}l}{A} +
h_{44}|_{\bar{b}}\frac{T'l}{2A}(A'^{2} + 1),\\ 
\delta n_{i} &=& 0\quad . 
\eearst

\subsection{Perturbation of the Junction Conditions}
With the complete normal to the brane in the perturbed background we can
calculate the perturbed second fundamental form on the brane. For the
spatial part we have 
\bear
K_{ij} &=& \bar{K}_{ij}  - \frac{\partial \delta n_{i} }{\partial x^{j}}
+ \bar{\Gamma}^{C}_{ij}|_{\bar{b}}\delta n_{C}  + \delta
\Gamma^{C}_{ij}|_{\bar{b}}\bar{n}_{C}\quad .
\eear

Using the bulk connections presented in the appendix  and the results
obtained for the perturbation of the normal, and using also $\bar{K}_{ij} =
T'\delta_{ij}l/A^2$, we find for the contribution of the perturbation
of the bulk to the second fundamental form on the brane
\bear
\delta K^{i}_{j} &=&
\delta^{i}_{j}\Bigl(h_{00}|_{\bar{b}}\frac{A'^{2}T'}{2l} +
h_{04}|_{\bar{b}}\frac{A'}{l}(A'^2 - 1) +
h_{44}|_{\bar{b}}\frac{T'}{2l}(A'^{2} - 1)\Bigr) \nonumber \\ 
  &+&\frac{A'A}{l}\Bigl[\frac{1}{2}(\frac{l^2}{A^2}\partial^{i}h_{0j}
|_{\bar{b}} + \partial_{j}h^{i}_{0}|_{\bar{b}}  -
\partial_{0}h^{i}_{j}|_{\bar{b}}) \Bigr]  \nonumber \\  
&+& \frac{T'A}{l}
\Bigl[\frac{1}{2}(\frac{l^{2}}{A^{2}}\partial^{i}h_{4j}|_{\bar{b}} +
\partial_{j}h_{4}^{i}|_{\bar{b}} - \partial_{4}h_{j}^{i}|_{\bar{b}})
\Bigr] \quad . \label{Kij1}
\eear

\subsubsection{Einstein Equations}
From the framework above developed, we learned that the general
perturbations in the bulk metric are translated to perturbations in the
brane metric. The junction conditions, now perturbed, relate the evolution
of the perturbation of the matter on the brane to the metric perturbations
introduced. However, we still have to solve the Einstein equations in the
bulk. Before doing that, it is useful to observe that there are five
reparametrization in the bulk coordinates that could be used to eliminate
five degrees of freedom in the perturbed metric. Thus, we fix the gauge in
the bulk by 
\be
h_{A4} = 0\quad .
\ee
With this choice we present the results for the perturbation in the Ricci
tensor in the bulk, 
\be
\delta R_{AB} = \partial_{C}\delta\Gamma^{C}_{AB} -
\partial_{A}\delta\Gamma^{C}_{BC} +
\bar{\Gamma}^{C}_{AB}\delta\Gamma_{CE}^{E} + \delta
\Gamma^{C}_{AB}\bar{\Gamma}_{CE}^{E} -
\bar{\Gamma}^{E}_{AC}\delta\Gamma^{C}_{EB} -
\delta\Gamma^{E}_{AC}\bar{\Gamma}^{C}_{EB}\quad .
\ee

The Einstein equations in the bulk read
\bear
\bar{R}_{AB} &=& \Lambda \frac{2l^{2}}{3(X^{4})^{2}}\eta_{AB}= -
\frac{4}{(X^{4})^{2}}h_{AB}\quad ,\\ 
\delta R_{AB} &=& \Lambda \frac{2l^{2}}{3(X^{4})^{2}}h_{AB}\quad ,
\eear
where we used the equations for the background metric.

We realize that the Einstein equations can be solved easily for a
transverse traceless-free perturbation when $\partial_{\mu}h^{\mu}_{\nu}
= 0= h^{\mu}_{\mu} $. Then components (4,0) and (4,4) are
automatically satisfied, remaining, for the five independent modes
$h_{\alpha\beta}$, 
\be
\frac{l^2}{(X^{4})^{2}}\partial^{A}\partial_{A}h_{\alpha\beta}  -
\frac{3}{X^{4}}\partial_{4}h_{\alpha\beta} =0\quad . 
\ee

Decomposing  $ h_{\alpha\beta} $ in Fourier modes in the variables $X^0$
and $X^i$, 
\[
h_{\alpha\beta} = \int d^{3}k dk^{0}
\tilde{\epsilon}_{\alpha\beta}(\vec{k},k^{0})F(X^{4})e^{ik^{0}X^{0} + i
\vec{k}.\vec{X}}\quad ,
\]
we find the following differential equation for the transformed function
$F(X^{4}) $,
\be
\frac{d^{2}}{d(X^{4})^{2}}F(X^{4}) -
\frac{3}{X^{4}}\frac{d}{dX^{4}}F(X^{4}) + \Bigl((k^{0})^{2} -
\mid\vec{k}\mid^{2}\Bigr)F(X^{4})= 0\quad .
\label{eBessel}
\ee

This  is just the Bessel differential equation of order 2. We write $F(X^{4}) 
= (mX^{4})^{2}Z(mX^{4})$, with $m^{2} = (k^{0})^{2} -\mid\vec{k}\mid^{2}$, 
in which case the general solution of the above equation is a combination
of Bessel or Hankel functions of order 2, 
\bearst
Z(mX^{4}) &=& Z^{(j)}_{2}(mX^{4}) = j_{m}J_{2}(mX^{4}) + n_{m}N_{2}(mX^{4})\\
          &=& Z^{(h)}_{2}(mX^{4}) = h^{(1)}_{m}H^{(1)}_{2}(mX^{4}) +
h^{(2)}_{m}H^{(2)}_{2}(mX^{4})\quad .
\eearst

Thus we have been able to solve exactly the bulk's perturbation
$h_{\alpha\beta}$ in the form 
\be
h_{\alpha\beta} = \int d^{3}k dm (mX^{4})^{2}
\epsilon_{\alpha\beta}(\vec{k},m)Z_{2}(mX^{4})e^{ik^{0}X^{0} + i
  \vec{k}.\vec{X}}\quad ,
\label{solbulk}
\ee
where $k^{\alpha}e_{\alpha\beta} = e^{\alpha}_{\alpha} = 0$
and $ (k^{0})^{2} = m^{2} + \mid\vec{k}\mid^{2}$.

\subsection{Perturbing the Brane Position}
In the unperturbed formalism we interpret the scale factor of the
Robertson-Walker brane as the position of the brane in the extra
dimension. If we introduce perturbations in the matter inside the brane,
there will be a perturbed Friedmann equation and, consequently, a
perturbation on the position of the brane. This brane bending effect
certainly may appear in the context of a reparametrization of the bulk,
showing that bulk metric perturbations and brane perturbations are related.

We work here in an unperturbed bulk, but with a perturbation in the
position of the brane in the extra dimension, \cite{Deruelle},  
\[
X^{A}(\eta,x^{i})|_{b} = \bar{X}^{A}(\eta) + \zeta(\eta,x^{i})\bar{n}^{A}.
\]

Therefore, the metric induced on the brane reads
\bear
ds^{2}_{b} = \frac{l^{2}}{A^{2}}\Bigr[\eta_{\mu\nu} - 2\frac{\zeta
A^{2}}{l^{2}}\bar{K}_{\mu\nu}\Bigl]dx^{\mu}dx^{\nu}\quad . 
\eear

Thus, a perturbation in the brane position induces a perturbation in the
brane metric, $\gamma_{\mu\nu}^{(p)} l^{2}/A^{2}$, where
\bear
\gamma^{(p)}_{\eta\eta} &=& -\frac{2\zeta}{l\sqrt{1+A'^{2}}} \Bigl(AA'' -
A'^{2} - 1\Bigr)\quad , \\ 
\gamma^{(p)}_{ij} &=& -\frac{2\zeta}{l}\sqrt{1+A'^{2}}\delta_{ij}\quad .
\eear

The junction conditions thus get perturbed, and we arrive at the result
\bear
K^{i}_{j} &=& \bar{K}^{i}_{j} + 
\partial_{j}\partial^{i}\zeta +  \frac{A^2}{l^2}\delta^{i}_{j}\Bigl[
\zeta\frac{A'^2}{A^2} + \frac{A'}{A}\zeta'  
\Bigr] \quad ,\nonumber \\
\delta K^{i}_{j} &=& \partial_{j}\partial^{i}\zeta +
\frac{A^2}{l^2}\delta^{i}_{j}\Bigl[ \zeta\frac{A'^2}{A^2} +
\frac{A'}{A}\zeta' \Bigr]\quad .   \label{Kij2}
\eear

\subsection{Gathering all together}
Gathering all results together, we conclude that
\bear
ds^2 &=& \frac{l^2}{(X^4)^2}\Bigl(\eta_{AB} + h_{AB}\Bigr)dX^A dX^B
\quad ,\nonumber \\
X^A &=& \bar{X}^A + \zeta \bar{n}^A\quad ,\nonumber\\
ds^2_b &=& a^2(\tau)\Bigl(\eta_{\mu\nu} + \gamma_{\mu\nu}\Bigr)dx^{\mu}
dx^{\nu} \quad , \label{unusualbmetric}
\eear
with
\bear
\gamma_{\eta\eta} &=& (1+H^2l^2)h_{00}|_{\bar{b}}
+\frac{2l\zeta}{\sqrt{1+H^2l^2}} \Bigl(\frac{\ddot{a}}{a} +
\frac{1}{l^2}\Bigr)\quad , \\ 
\gamma_{\eta i} &=& \sqrt{1+H^2l^2}h_{0i}|_{\bar{b}}\quad , \\
\gamma_{ij} &=&
h_{ij}|_{\bar{b}}-\frac{2\zeta}{l}\sqrt{1+H^2l^2}\delta_{ij}\quad .
\eear

The dependence in the bulk coordinates  of the five degrees of freedom
associated with the perturbation of the bulk metric, $h_{\alpha\beta}$, is
known from the Einstein equations in the bulk. What remains to be done is
to impose the boundary conditions on the general solution and to relate
those fields with the matter perturbation on the brane. The junction
conditions provide those relations between the perturbations. Here, however,
as comparing with \cite{Deruelle}, we do not look for the equations
restricted to a specific brane evolution, but make a full treatment of the
junction conditions leading to the full equations of motion on the brane
for a generic dynamics. 

The unperturbed junction conditions give the usual conservation of
energy-momentum on the brane and the evolution of the scale factor by a
Friedmann equation. On the other hand, if we allow  matter perturbations
to exist on the brane, as the bulk energy continues to be just a
cosmological constant, the perturbed part of the junction conditions
implies 
\bear
\delta\Bigr({\cal T}^{\mu}_{\nu ;\mu}\Bigl) &=& 0\quad ,\nonumber \\
\delta K^i_j &=& \frac{\kappa}{2}\Bigl(\delta {\cal T}^i_j -
\frac{1}{3}\delta^i_j \delta{\cal T}\Bigr)\quad .
\eear
Using the previous results we find the relations
\bear
&&\delta {\cal T}^{j}_{\eta} = -\delta{\cal T}_{j}^{\eta} +
\gamma_{\eta\beta}\bar{{\cal T}}_{j}^{\beta} + \gamma^{j\alpha}\bar{{\cal
T}}_{\alpha}^{\eta} = -\delta{\cal T}_{j}^{\eta} + \gamma_{\eta}^{j}(P +
\rho)\quad , \\ 
&&\dot{\delta{\cal T}^{\eta}_{\eta}} + \frac{1}{a}\partial_{i}\delta{\cal
T}^{i}_{\eta}  +3H\delta{\cal T}^{\eta}_{\eta}- H \delta{\cal T}^{k}_{k} -
\frac{\rho+P}{2}\dot{\gamma^{k}_{k}} = 0\quad ,\\
&&\dot{\delta{\cal T}^{\eta}_{i}} + \frac{1}{a}\partial_{j}\delta{\cal
T}^{j}_{i}   +4H \delta{\cal T}^{\eta}_{i}   +
\frac{P+\rho}{2a}\partial_{i}\gamma^{\eta}_{\eta} = 0\quad ,
\eear
where $(\:\dot{}\:) = \partial /\partial \tau$.

In order to put the equations in a familiar form, we parametrize the
perturbations by the functions $\delta \rho$, $\delta P$, the velocity
$v_{i}$ and the anisotropic stress $\Sigma^{i}_{j}$ with the definitions
\bear
\delta{\cal T}^{\eta}_{\eta} &=& -\delta \rho\quad , \\
\delta{\cal T}^{i}_{\eta} &=& -(\rho + P)v^{i}\quad , \\
\delta{\cal T}^{\eta}_{i} &=& (\rho + P)(v_{i} - \gamma^{\eta}_{i} )\quad , \\
\delta{\cal T}^{i}_{j} &=& \delta P\delta^{i}_{j} + \Sigma^{i}_{j}\quad .
\eear
As a consequence we find, using $P = \omega \rho$, 
\bear
&&\dot{\delta\rho} + \frac{\rho(1+\omega)}{a}\partial_{i}v^{i} +
3H(\delta\rho+\delta P) + \frac{\rho(1+\omega)}{2}\dot{\gamma^{k}_{k}} = 0
\label{juncons0}\\
 (1+\omega)\dot{v}_{i} &+& \frac{1}{a\rho}\partial_{j}\Bigl(\delta
P\delta^{j}_{i} + \Sigma^{j}_{i}\Bigr) + \Bigl[H(1- 3\omega)(1+\omega) +
\dot{\omega}\Bigr](v_{i} - \gamma^{\eta}_{i} ) \nonumber
\\ &+& (1+\omega)\Bigl(\frac{1}{2a}\partial_{i}\gamma^{\eta}_{\eta} -
\dot{\gamma}_{i}^{\eta}\Bigr) = 0\quad .  \label{junconsk}
\eear

The expressions above determine the dynamics of the perturbations on the
brane. Once we solve the junction conditions 
and express $\zeta$ in terms of $h_{\alpha\beta}$,
we will be able to find the expressions for $\gamma_{\mu\nu}$ and, using
(\ref{juncons0}) and (\ref{junconsk}), finally study the evolution of
matter perturbations on the brane. 

We first decouple the equations in scalar, vector and tensorial
modes, writting, in Fourier space,
\bear
v_{i} &=& -i\frac{k_i}{k}V + v_{i}^{V}, \label{energy} \\
\Sigma_{ij} &=& \Bigl(-\frac{k_{i}k_{j}}{k^2} +
\frac{1}{3}\delta_{ij}\Bigr)\Sigma  - \frac{i}{2k}(k_{i}\Sigma_{j} +
k_{j}\Sigma_{i}) + \Sigma^{T}_{ij}\quad , \label{stress}
\eear
where $k^{i}\Sigma^{T}_{ij} = k^{i}\Sigma_{i} = 0$. 

We now choose $k^{i} = (0,0,k)$ and find 
\bear
 (1+\omega)\dot{V} &-& \frac{k}{a\rho}\Bigl(\delta
P -\frac{2}{3}\Sigma\Bigr) + \Bigl[H(1- 3\omega)(1+\omega) +
\dot{\omega}\Bigr](V  - i \gamma^{\eta}_{3} ) \nonumber \\ 
& +& (1+\omega)\Bigl(-\frac{k}{2a}\gamma^{\eta}_{\eta} -i
\dot{\gamma}_{3}^{\eta}\Bigr) = 0 \label{junconskdeco1}
\eear
for the scalar part and
\bear
 (1+\omega)\dot{v}^{V}_{i} &+& \frac{k}{2a\rho}\Sigma_{i} + \Bigl[H(1-
3\omega)(1+\omega) + \dot{\omega}\Bigr]( v_{i}^{V} - \gamma^{\eta}_{i} )
\nonumber \\ 
&-& (1+\omega)\dot{\gamma}_{i}^{\eta} = 0 \label{junconskdeco2}
\eear
for the vector part, where $i = 1,2$.

\subsubsection*{Junction Conditions}
Following \cite{Deruelle}, we define the quantity
\bear
F^{i}_{j} &=& - \frac{\kappa}{2}\Sigma^{i}_{j} +
\partial_{j}\partial^{i}\zeta \nonumber\\
&-&
\frac{Hl}{2a}\Bigl[(a^2\partial^{i}h_{0j}|_{\bar{b}} +
\partial_{j}h^{i}_{0}|_{\bar{b}}  - \partial_{0}h^{i}_{j}|_{\bar{b}})
\Bigr] - \frac{\sqrt{1+H^2l^2}}{2a}
\partial_{4}h_{j}^{i}|_{\bar{b}} \quad . \label{Fij} 
\eear
We find that this quantity has no traceless part, that is
$F^{i}_{j} = \frac{1}{3}\delta^{i}_{j}F$, and the trace reads
\bear
F = \frac{\kappa}{2}\delta \rho - \Bigl[ 3\zeta H^2 - 3H\dot{\zeta} +
3h_{00}|_{\bar{b}}\frac{H^2l}{2}\sqrt{1+H^2l^2}\Bigr]\quad .
\eear

If we choose the spatial coordinates such that $k^{i} = \delta^{i}_{3}k$,
the traceless-free condition implies that 
\bear
\epsilon_{0i} &=& -\frac{k}{k^{0}}\epsilon_{3i} = -\frac{k}{\sqrt{m^2 +
k^2}}\epsilon_{3i} \quad , \label{TTfree0} \\ 
\epsilon_{00} &=& \frac{k^2}{(k^{0})^2}\epsilon_{33} = \frac{k^2}{m^2 +
k^2}\epsilon_{33} \quad , \\ 
\epsilon_{22} &=& - \epsilon_{11} - \frac{m^2}{m^2 + k^2}\epsilon_{33}
\quad .  \label{TTfree}
\eear
Only five degrees of freedom remain, and they are $\epsilon_{12}$,
$\epsilon_{13}$, $\epsilon_{23}$, $\epsilon_{33}$ and $\epsilon_{11}$.  

A thorough discussion of massless and massive modes is given in the
appendices.

\subsection{Connecting with the usual cosmological perturbation theory}

In order to clarify what is hidden behind those equations, let us take
a closer look into the usual cosmological perturbation theory. The
standard approach is to perturb the RW metric in  the form
\be
ds^{2} = a^2(\eta)\Bigl(-(1+2A)d\eta^{2} - B_{i}d\eta dx^{i} +
[(1+2D)\delta_{ij} + E_{ij}]dx^{i}dx^{j}\Bigr) \quad , \label{usualbmetric}
\ee
with $E_{ij}$ traceless.

Then we define the quantities $B_{i}^{V}$, $E_{i}$ and
$E_{ij}^{T}$ such that \cite{Liddle}
\bear
B_{i} &=& -\frac{ik_{i}}{k}B + B_{i}^{V} \\
E_{ij} &=& \Bigl(-\frac{k_{i}k_{j}}{k^{2}} + \frac{1}{3}\delta_{ij}\Bigr)E
-\frac{i}{2k}\Bigl(k_{i}E_{j} + k_{j}E_{i}\Bigr) + E_{ij}^{T}\quad ,
\label{decomposition}
\eear
with $k^{i}B_{i}^{V} = k^{i}E_{i} = k^{i}E_{ij}^{T} = 0$. This general
decomposition of the metric perturbations, together with
(\ref{energy}) and (\ref{stress}) decouple the equations between what
is called scalar perturbations ($\delta \rho$, $\delta P$, $V$,
$\Sigma$, $A$, $B$, $D$, $E$), vectorial perturbations ($v_{i}^{V}$,
$\Sigma_{i}$, $B_{i}^{V}$, $E_{i}$) and tensorial perturbations
($\Sigma_{ij}^{T}$, $E_{ij}^{T}$).

The  gauge invariant  tensorial perturbations in usual cosmology satisfy
the wave equation 
\be
(E_{ij}^{T})'' + 2aH(E_{ij}^{T})'+ k^{2}E_{ij}^{T} = 8\pi G a^2
P\Sigma_{ij}^{T}\quad .
\ee

For the scalar and vector sectors, we must fix the gauge and the 
 field equations, give the evolution of the perturbations and some
constraint equations. For example, the scalar sector in the
longitudinal gauge ($B=E=0$) reads  \cite{Liddle}
\be
\frac{\delta \dot{\rho}}{\rho} = -(1+\omega)\frac{k}{a}V - 3H\Bigl(\frac{\delta
\rho}{\rho} + \frac{\delta P}{\rho}\Bigr) - 3(1+\omega)\dot{D} \quad ,
\label{eq1usual}
\ee
and
\be
 \dot{V} = -H(1-3\omega)V - \frac{\dot{\omega}}{(1+\omega)}V +
\frac{k}{a(\rho + P)}\Bigl(\delta P - \frac{2}{3}\Sigma\Bigr) +
\frac{k}{a}A\quad . \label{eq2usual}
\ee
The constraint equations relate directly the metric perturbation to
the matter perturbation. In this gauge, for the scalar sector, we find
\be
k^2D = 4\pi Ga^2\Bigl(\delta \rho + \frac{3aH}{k}(\rho+P)V\Bigr)
\label{con1usual}
\ee
and
\be
k^2(A+D) = -8\pi G a^{2}\Sigma \quad .
\label{con2usual}
\ee

Now, returning to the brane world cosmology, we can find a connection
between our treatment and the usual cosmology from the perturbed brane
metric. Comparing (\ref{unusualbmetric}) with (\ref{usualbmetric}) and 
separating the trace part of $h_{ij}$ ($h_{00} = h_{11} + h_{22} + h_{33}$)
and of $E_{ij}$ ($E_{11} + E_{22} = -E_{33}$) we note that
\bear
-2A &=& \gamma_{\eta\eta}  = (1+H^2l^2)h_{00}|_{\bar{b}} +
\frac{2l\zeta}{\sqrt{1+H^{2}l^{2}}}\Bigl(\frac{\ddot{a}}{a} +
\frac{1}{l^2}\Bigr) \quad ,\\
-B_{i} &=& 2\gamma_{\eta i} = 2\sqrt{1+H^{2}l^{2}}h_{0i} \quad , \\
2D &=& \frac{1}{3}\gamma^{k}_{k}  =  \frac{1}{3}h_{00} -
\frac{2\zeta}{l}\sqrt{1+H^{2}l^2} \quad ,\\ 
2E_{ij} &=& \gamma_{ij} - \frac{1}{3}\gamma^{k}_{k}\delta_{ij} \quad  .
\eear

Explicitly the last expression reads
\bear
2E_{11} &=&\frac{2}{3}h_{11}- \frac{1}{3}h_{22} - \frac{1}{3} h_{33} \quad , \\
2E_{22} &=&\frac{2}{3}h_{22}-\frac{1}{3}h_{11} - \frac{1}{3} h_{33} \quad ,\\
2E_{33} &=&\frac{2}{3}h_{33} - \frac{1}{3}h_{11} - \frac{1}{3} h_{22} \quad ,\\
2E_{12} &=&h_{12} \quad , \\
2E_{13} &=&h_{13} \quad , \\
2E_{23} &=&h_{23}  \quad ,
\eear
and we have the correct number of functions to
describe perturbations on the brane.

Now, for $k^{i} = (0,0,k)$, the scalar - vector - tensorial decomposition
(\ref{decomposition}) states the two polarizations of tensorial
perturbations as described by
\bear
E_{11}^{T} &=&\frac{1}{4}(h_{11}- h_{22}) \quad ,\label{einicio} \\
E_{12}^{T} &=& \frac{1}{2}h_{12} \quad .
\eear
This leads us to a direct connection with standard perturbation theory. We
shall find that the real tensorial modes are the 11 and 22 modes, but not
the 13 or 23 ones.

The vectorial components are
\bear
B_{1}^{V} &=& -2\gamma_{\eta 1} = -2\sqrt{1+H^{2}l^{2}}h_{01} \quad , \\
B_{2}^{V} &=& -2\gamma_{\eta 2} = -2\sqrt{1+H^{2}l^{2}}h_{02} \quad , \\
E_{1} &=& ih_{13} \quad , \\
E_{2} &=& ih_{23} \quad ,
\eear
and finally, the scalar ones are
\bear
A &=&-\frac{1}{2}\gamma_{\eta\eta} =  -\frac{1}{2}(1+H^2l^2)h_{00}|_{\bar{b}} -
\frac{l\zeta}{\sqrt{1+H^{2}l^{2}}}\Bigl(\frac{\ddot{a}}{a} +
\frac{1}{l^2}\Bigr) \quad ,\\
D &=& -\frac{1}{6}\gamma^{k}_{k}   = -\frac{1}{6}h_{00} +
\frac{\zeta}{l}\sqrt{1+H^{2}l^2} \quad ,\\ 
E &=& \frac{1}{4}h_{00} - \frac{3}{4}h_{33} \quad ,\\
B &=& -2i\gamma_{\eta 3} =  -2i\sqrt{1+H^{2}l^{2}}h_{03}\quad .
\label{efim}
\eear 
 
The general metric perturbations are described in usual cosmology by 10
functions ($A$, $B$, $D$, $E$, $B_{1}$, $B_{2}$, $E_{1}$, $E_{2}$, $E^{T}_{11}$
and $E^{T}_{12}$) matching  the correct number of degrees of freedom given 
by the traceless symmetric part of $h_{\mu\nu}$ and the field $\zeta$. 
However, when we require $k^{\mu}h_{\mu\nu} = 0$, we force the existence of 
four constraint equations that reduce this number to 6, which  is the
number of independent perturbations after the gauge choice in usual
cosmology. Thus, from the point of view of the brane, we completely
fixed the gauge when we demand $k^{\mu}h_{\mu\nu} = 0$.  This is an unusual 
gauge choice, it is neither a longitudinal (Newtonian) gauge nor
a total-matter gauge. It relates the scalar perturbations with each
other without canceling any one of them.

We can see that the brane-world treatment described so far walks
closely to the usual cosmological perturbation theory. In fact, from the
analysis developed in the last section, we also find that dynamical equations
are quite similar to (\ref{eq1usual}) and (\ref{eq2usual}). In terms of
$B$, $D$, and $A$, the scalar sector,  (\ref{juncons0}) and
(\ref{junconskdeco1}) read
\bear
\frac{\dot{\delta{\rho}}}{\rho} + (1+\omega)\frac{k}{a}{V} +
3H(\frac{\delta{\rho}}{\rho}+\frac{\delta {P}}{\rho}) +
(1+\omega)3\dot{D} = 0\quad ,
\eear
and
\bear
 \dot{V} =  \frac{k}{a(\rho + P)}\Bigl(\delta
P -\frac{2}{3}\Sigma\Bigr) - \Bigl[H(1- 3\omega) +
\frac{\dot{\omega}}{1+\omega}\Bigr](V  + \frac{B}{2} ) +
\Bigl(\frac{k}{a}A -  \frac{1}{2}\dot{B}\Bigr)\quad . \nonumber \\
\eear

Comparing with (\ref{eq1usual}) and (\ref{eq2usual}) we see that the only
difference comes from the unusual gauge choice we have 
made (in the Newtonian gauge $B=0$ and this expressions are identical
to the usual ones).

On the other hand, the quite different behaviour of the brane-world
perturbations comes from the constraint equations. Contrary to the
usual cosmological scenario there are constraints between tensorial
modes and the anisotropic stress on the brane. Denoting the quantities under a hat by the fourier modes as in
\bear
h_{\alpha\beta} &=& \int d^{3}k dm \:\hat{h}_{\alpha\beta}(m,k;X^{0},X^{4})\:
e^{i\vec{k}.\vec{x}}\quad 
\eear
and recording the dependence of $\hat{h}$ on $X^{0}$ and $X^{4}$, (\ref{solbulk}), we follow the procedure described in the appendix for massive and massless modes and use the connection with usual cosmology theory, (\ref{einicio})-(\ref{efim}), to find for the tensorial sector
\bear
\frac{\kappa}{2a}\hat{\Sigma}^{T}_{12} &=&  -
\hat{E}_{12}^{T}\Bigl[iHl\sqrt{m^2 + k^2} + m\sqrt{1+H^2l^2}
\frac{Z_1(mla^{-1}(\tau))}{Z_2(mla^{-1}(\tau))}\Bigr]\quad , \label{grav12} \\
 \frac{\kappa}{2a}\hat{\Sigma}^{T}_{11} &=& -
\hat{E}_{11}^{(T)}\Bigl[iHl\sqrt{m^2 + k^2} + m\sqrt{1+H^2l^2}
\frac{Z_1(mla^{-1}(\tau))}{Z_2(mla^{-1}(\tau))}\Bigr]\quad , \label{grav11}
\eear
for the vectorial sector
\bear
\frac{\kappa}{2a}\hat{\Sigma}_{1} &=&
-\hat{E}_{1}\Bigl[\frac{iHlm^{2}}{\sqrt{k^2+m^2}}+
m\sqrt{1+H^2l^2}\frac{Z_1(mla^{-1}(\tau))}{Z_2(mla^{-1}(\tau))}\Bigr]\quad , \\ 
\frac{\kappa}{2a}\hat{\Sigma}_{2} &=&  -\hat{E}_{2}\Bigl[\frac{iHl
m^2}{\sqrt{k^2+m^2}}   +
m\sqrt{1+H^2l^2}\frac{Z_1(mla^{-1}(\tau))}{Z_2(mla^{-1}(\tau))}\Bigr]  \quad ,
\eear
and for the scalar sector
\bear
F  &=&  -
\frac{\hat{h}}{2a}\epsilon_{00}\Bigl[-iHl\sqrt{m^2
+ k^2} + m\sqrt{1+H^2l^2}
\frac{Z_1(mla^{-1}(\tau))}{Z_2(mla^{-1}(\tau))}\Bigr]  -
\frac{k^2}{a^2}\hat{\zeta}\quad , \nonumber \\ 
 \frac{\kappa}{a^2}\hat{\Sigma} &=& -
 \frac{2}{a}\hat{E}\Bigl[iHl\sqrt{m^2 + k^2} + m\sqrt{1+H^2l^2}
\frac{Z_1(mla^{-1}(\tau))}{Z_2(mla^{-1}(\tau))}\Bigr]\nonumber \\ &
-&4iHl\sqrt{k^{2}+m^{2}}\frac{\hat{h}}{2a}\epsilon_{00}  +
2\frac{k^2}{a^2}\hat{\zeta}\quad . 
\eear

These are our main results: we completely solve the wave equations in the bulk and found the junction conditions for arbitrary brane dynamics in each mode sector.
Our equations are quite complicated because of the effect of massive 
modes and the unusual gauge choice we made. We should indeed expect constraints in the scalar and in the vector sectors. 
However equations similar to 
(\ref{grav12}) and (\ref{grav11}) do not appear in usual cosmology. When we 
are talking about a Minkowsky brane with $a = a_{0}$, we can avoid such 
constraints choosing $m$ such that $Z_{1}(ml/a_{0}) =0$. Indeed this is 
exactly what is done in the normal  coordinate approach \cite{Easther}. 
However, in normal coordinates, we can not treat the full problem
analytically in generic dynamical universes. From the point of view of the
bulk, this seems to be possible and, after fixing a gauge in an unusual
form, we are able to solve analytically the bulk modes, finding the 
constraints equations for general brane behaviour. As tensorial modes are
gauge independent, from our analysis, we can conclude that brane-world
cosmology predicts unusual constraint equations between tensorial modes
and the stress energy tensor. 

In the usual cosmology, gravitational waves ($E_{11}^{T}$ and $E_{12}
^{T}$) couple with the tensorial part of the anisotropic stress, but 
there are no algebraic constraint between them. Here we see that, besides special cases 
discussed in the next section, in the great majority of possibilities of brane dynamics, there 
are no gravitational waves without anisotropic stress on the brane. 
Clearly any unusual behaviour as this one comes from the fact that
gravitational waves should be fundamentally affected in our model. In 
a generic brane world scenario, we cannot talk about a gravitational wave 
strictly restricted to the brane. The extra constraints appear because 
strictly speaking, gravitational waves are not brane objects. They 
belong to the bulk and so they must satisfy the bulk Einstein equations. 
As they travel through the bulk, we can expect to find waves that 
follow closely the brane evolution and even go across it. For brane
world residents, the amplitude of what is usually called a gravitational 
wave is, in fact, an average over all amplitudes of the bulk waves that, 
at a certain moment, cross the brane. Because the junction conditions 
force the matter content on the brane to behave in correspondence to the 
bulk geometry,  all possible tensorial perturbations calculated on the 
brane appear to brane world residents as anisotropic stress on the 
energy-momentum tensor.

\section{Normal Coordinates}

During a de-Sitter expansion, we can explicitly transform the bulk
coordinates ($X^0$, $X^4$) to normal coordinates ($t$, $y$) where the
background metric is written by
\bear
ds^{2} = \Bigl(\cosh(y/l) - \sqrt{1+H^2l^2}\sinh(y/l)\Bigr)\Bigl(-dt^2
+ a^{2}(t)\delta_{ij}dx^i dx^j\Bigr) + dy^{2}
\eear
with $0 \le y < \tanh^{-1}\Bigl[(1+H^2l^2)^{-1/2}\Bigr]$ and the brane
position is taken at $y=0$.

We note that the relevant transformation is
\bear
X^{4} = X^{4}(t,y) = \frac{l}{a(t)[\cosh(y/l) -
\sqrt{1+H^2l^2}\sinh(y/l)]}\quad , \\
X^{0} = X^{0}(t,y) = \frac{1}{Ha(t)}\sqrt{1 + \frac{H^2l^2}{\cosh(y/l) -
\sqrt{1+H^2l^2}\sinh(y/l)]}}\quad . \\
\eear

Thus, we can write the solution for pure tensor modes with respect to
the brane metric found in the previous sections in normal
coordinates. Writing
\be
h_{ij}(t,\vec{x}, y) = \int d^{3}k dm \epsilon_{ij}(\vec{k},m)
\hat{h}_{km}(t,\vec{x},y)\quad , 
\ee
we find for a bounded zero mode
\bear
\hat{h}_{km}(t,\vec{x},y) &=&
\exp{[i\vec{k}.\vec{x}]}\frac{m^2l^2}{a(t)^2[\cosh(y/l) -
\sqrt{1+H^2l^2}\sinh(y/l)]^2}\times\nonumber \\ 
&&N_{2}\Bigl(\frac{ml}{a(t)[\cosh(y/l) -
\sqrt{1+H^2l^2}\sinh(y/l)]}\Bigr) \times \nonumber \\ &&
\exp{\Bigl[\frac{i\sqrt{m^2 + 
\mid\vec{k}\mid^2}}{Ha(t)}\sqrt{1 + \frac{H^2l^2}{\cosh(y/l) -
\sqrt{1+H^2l^2}\sinh(y/l)}}\Bigr]}\quad .\nonumber 
\eear

We can plot the above solutions with respect to the extra
coordinate $y$ and note the decaying behaviour of the perturbations for
massive modes away from the brane.

\begin{figure*}[htb!]
\begin{center}
\leavevmode
\begin{eqnarray}
\epsfxsize= 9.0truecm
{\epsfbox{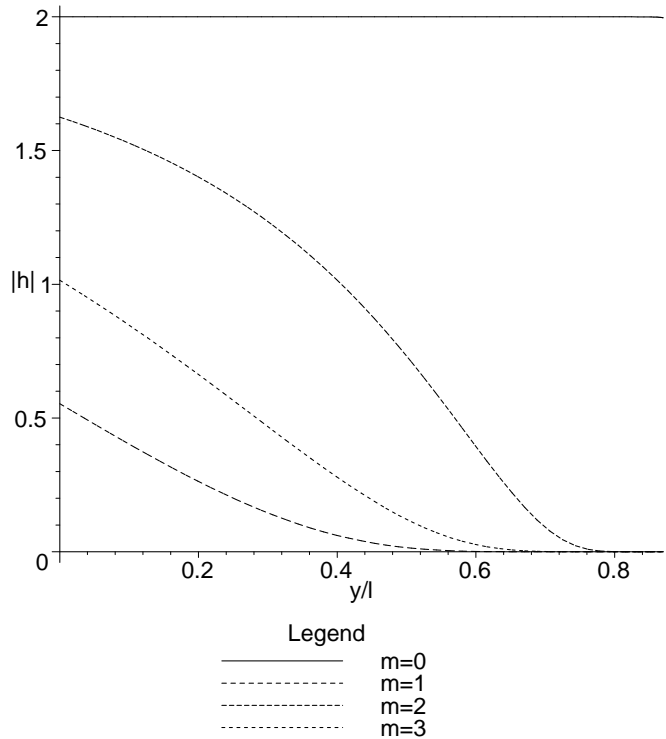}}\nonumber
\end{eqnarray}
\caption{Plot of $\mid h_{km}(t,\vec{x},y)\mid$ with respect to
$y$. The brane is in $y=0$ and we used $a(0)H = lH =1$. We ploted the
modes $m=0$, $1$, $2$, and $3$ in this order (from up to down in the
graphic), showing the
behaviour of massive amplitudes out of the brane.}
\label{graphprimitivo}
\end{center}
\end{figure*}

The junction conditions are reproduced in the absence of anisotropic
stress by 
\be
\frac{\partial  h(t,\vec{x},y) }{\partial y}\Bigr|_{y=0} = 0
\ee
This equation can not be solved non trivially for all times because the
solution just found is non-separable in $y$ and $t$.

In a de-Sitter background, it was shown that it is possible to
separate the wave equation in normal coordinates and find a separable
solution in $y$ and $t$,  \cite{Koyama}, \cite{Easther},
\cite{Frolov}, \cite{du}, with gravitational waves in the ausence of brane
anisotropic stress. However, for a non-de Sitter evolution, this 
is not possible and, because we started the approach generically, our
solution is intrinsically non-separable. Clearly, from this non
separability in the bulk approach, the connection with the usual de-Sitter
solution becomes obscure and, for this reason, we still need
a method to determine the initial conditions and choose the amplitudes
$\epsilon_{ij}(\vec{k},m)$ to finally solve the spectrum of
tensorial perturbations. However, 
because such non separability is inherent of generic brane dynamics, our
results have far reaching consequences for inflation: it gives us, for
instance, a mechanism for anisotropy creation in the brane-world. If
initial conditions, during a de-Sitter inflationary phase, created
tensorial perturbations propagating in the bulk with an anisotropic stress
free brane world, as suggested in usual inflationary models
\cite{Easther}, after inflation, and also during the reheating, those
waves, interacting with the brane, would necessarily produce tensorial
modes of anisotropy in our universe. Those anisotropic modes would be
created by the gravitational waves as described by the complete junctions
conditions found in the last section.

\appendix

\section{Notation.}
Unperturbed tangent vectors are denoted by a bar $\bar{V}^{A}_{\mu}$. 
Quantities calculated on the unperturbed brane are followed by
$|_{\bar{b}}$, while $|_{b}$ denotes the perturbed brane.  

Upper case latin indices parametrize the bulk, $A,B,C, = 0..4$ (4 is the
extra dimension), greek indices refers to the usual space-time, $\mu, \nu,
\lambda = 0..3$, and the spatial part in the brane is parametrized by
lower case latin indices,  $i,j,k = 1..3$. 
\section{Connections}
In this section it is presented for reference the non-null connections of
the unperturbed and perturbed geometry. 
\subsection{Brane Connections}
For the RW metric in conformally minkowskian coordinates,
\be
ds^2 = \frac{l^2}{A^2}\eta_{\mu\nu}dx^{\mu}dx^{\nu},
\ee
the non-zero connections are
\bearst
\bar{\Gamma}^{\eta}_{\eta\eta} &=& - \frac{A'(\eta)}{A}\quad,\qquad
\bar{\Gamma}^{\eta}_{ij} = - \frac{A'(\eta)}{A}\delta_{ij} \quad , \\
\bar{\Gamma}^{i}_{\eta j} &=& - \frac{A'(\eta)}{A}\delta^{i}_{j}\quad .
\eearst

Linearly in the perturbation, the non-zero perturbed connections
calculated with the perturbed RW metric $
ds^2 = \frac{l^2}{A^2}(\eta_{\mu\nu} + \gamma_{\mu \nu})dx^{\mu}dx^{\nu}$ are
\bear
\delta \Gamma^{\eta}_{\eta\eta} &=&
\frac{1}{2}\partial_{\eta}\gamma^{\eta}_{\eta} \quad , \\ 
\delta \Gamma^{\eta}_{\eta i} &=&
\frac{1}{2}\partial_{i}\gamma^{\eta}_{\eta} +
\frac{A'(\eta)}{A}\gamma^{\eta}_{i} \quad ,\\ 
\delta \Gamma^{\eta}_{i j} &=& \frac{1}{2}(\partial_{i}\gamma^{\eta}_{j} +
\partial_{j}\gamma^{\eta}_{i} + \partial_{\eta}\gamma_{ij}) -
\frac{A'(\eta)}{A}(\gamma_{ij} + \gamma_{\eta\eta}\delta_{ij}) \quad , \\ 
\delta \Gamma^{i}_{\eta \eta} &=& \partial_{\eta}\gamma^{i}_{\eta}
-\frac{1}{2} \partial_{i}\gamma^{\eta\eta}  -
\frac{A'(\eta)}{A}\gamma^{i}_{\eta} \quad , \\ 
\delta \Gamma^{i}_{j \eta} &=& \frac{1}{2}(\partial_{j}\gamma^{i}_{\eta} +
\partial_{\eta}\gamma^{i}_{j} - \partial_{i}\gamma_{j \eta}) \quad , \\ 
\delta \Gamma^{i}_{jk} &=& \frac{1}{2}(\partial_{j}\gamma^{i}_{k} +
\partial_{k}\gamma^{i}_{j} - \partial_{i}\gamma_{j k})
-\frac{A'(\eta)}{A}\delta_{jk}\gamma^{i\eta}\quad . 
\eear

It is worthwhile reminding that the indices of the perturbation in the
metric, $\gamma_{\mu\nu}$, are raised and lowered with
$\eta_{\mu\nu}$. The same holds for the perturbation of the bulk metric
described below, $h_{AB}$, which are raised and lowered by $\eta_{AB}$.

\subsection{Bulk Connections}
For the AdS bulk in conformally minkowskian coordinates described by the
metric $ds^{2} = \frac{l^{2}}{(X^{4})^{2}}\eta_{AB}dX^{A}dX^{B}$,
the non-zero connections read
\bearst
\bar{\Gamma}^{0}_{04} &=& \bar{\Gamma}^{4}_{00} = \bar{\Gamma}^{4}_{44} =
-\frac{1}{X^{4}}\quad ,\\ 
\bar{\Gamma}^{4}_{ij} &=& \frac{\delta_{ij}}{X^{4}}\quad ,\qquad
{\hbox{and}} \qquad
\bar{\Gamma}^{i}_{4j} = -\frac{\delta^{i}_{j}}{X^{4}} \quad .
\eearst

In the linear approxiamtion, the non-zero perturbations in the
connections computed with the perturbed AdS metric $
ds^2 = \frac{l^2}{(X^{4})^2}(\eta_{AB} + h_{AB})dX^{A}dX^{B}$, are
\bear
\delta \Gamma^{0}_{00} &=& \frac{1}{2}\partial_{0}h^{0}_{0} +
\frac{h^{04}}{X^{4}} \quad ,\\ 
\delta \Gamma^{0}_{40} &=& \frac{1}{2}\partial_{4}h^{0}_{0} \quad ,\\
\delta \Gamma^{0}_{44} &=& -\frac{l^{2}}{2(X^{4})^2}\partial^{0}h_{44} +
\partial_{4}h^{0}_{4} - \frac{h^{04}}{X^{4}} \quad ,\\ 
\delta \Gamma^{0}_{ij} &=& \frac{1}{2}(\partial_{i}h^{0}_{j} +
\partial_{j}h_{i}^{0}  - \frac{l^{2}}{(X^{4})^2}\partial^{0}h_{ij}) -
\frac{h^{04}}{X^{4}}\delta_{ij} \quad ,\\ 
\delta \Gamma^{0}_{i0} &=& \frac{1}{2}\partial_{i}h^{0}_{0} \quad ,\\
\delta \Gamma^{0}_{i4} &=& \frac{1}{2}(\partial_{i}h^{0}_{4} +
\partial_{4}h_{i}^{0} - \frac{l^{2}}{(X^{4})^2} \partial^{0}h_{i4}) \quad ,\\ 
\delta \Gamma^{4}_{00} &=& \partial_{0}h_{0}^{4} -
\frac{l^{2}}{2(X^{4})^2}\partial^{4}h_{00} + \frac{h^{44}}{X^{4}} +
\frac{h^{00}}{X^{4}} \quad ,\\ 
\delta \Gamma^{4}_{40} &=& \frac{1}{2}\partial_{0}h_{44} +
\frac{h_{04}}{X^{4}} \quad ,\\ 
\delta \Gamma^{4}_{44} &=& \frac{1}{2}\partial_{4}h_{4}^{4} \quad ,\\ 
\delta \Gamma^{4}_{i4} &=&  \frac{1}{2}\partial_{i}h_{44}
+\frac{h_{i}^{4}}{X^{4}} \quad ,\\ 
\delta \Gamma^{4}_{ij} &=&  \frac{1}{2}(\partial_{i}h_{4j} +
\partial_{j}h_{4i} - \frac{l^{2}}{(X^{4})^2}\partial^{4}h_{ij})
+\frac{h_{ij}}{X^{4}} - \frac{h^{4}_{4}}{X^{4}}\delta_{ij} \quad ,\\ 
\delta \Gamma^{i}_{00} &=& - \frac{l^{2}}{2(X^{4})^2}\partial^{i}h_{00}
+\frac{h^{i}_{4}}{X^{4}} + \partial_{0}h_{0}^{i} \quad ,\\ 
\delta \Gamma^{i}_{04} &=&  \frac{1}{2}(\partial_{0}h_{4}^{i} +
\partial_{4}h_{0}^{i} - \frac{l^{2}}{(X^{4})^2}\partial^{i}h_{04}) \quad ,\\ 
\delta \Gamma^{i}_{44} &=& \partial_{4}h_{4}^{i} -\frac{h^{i}_{4}}{X^{4}}
- \frac{l^{2}}{2(X^{4})^2}\partial^{i}h_{44} \quad ,\\ 
\delta \Gamma^{i}_{0j} &=&  \frac{1}{2}(\partial_{j}h^{i}_{0} +
\partial_{0}h^{i}_{j} - \frac{l^{2}}{(X^{4})^2}\partial^{i}h_{0j}) \quad ,\\ 
\delta \Gamma^{i}_{4j} &=&  \frac{1}{2}(\partial_{j}h^{i}_{4} +
\partial_{4}h^{i}_{j} - \frac{l^{2}}{(X^{4})^2}\partial^{i}h_{4j}) \quad , \\ 
\delta \Gamma^{i}_{kj} &=&  \frac{1}{2}(\partial_{j}h^{i}_{k} +
\partial_{k}h^{i}_{j} - \frac{l^{2}}{(X^{4})^2}\partial^{i}h_{kj}) -
\frac{h^{i}_{4}}{X^{4}}\delta_{kj} \quad . 
\eear


\section{Massive Modes}
Let us write
\bear
h_{\alpha\beta} &=& \int d^{3}k dm \:\hat{h}(m,k;X^{0},X^{4})\:
\epsilon_{\alpha\beta}(m,k)e^{i\vec{k}.\vec{x}}\quad , \\
h_{\alpha\beta}|_{\bar{b}} &=& \int d^{3}k dm \:\hat{h}(m,k;\tau)\:
\epsilon_{\alpha\beta}(m,k)e^{i\vec{k}.\vec{x}} , \\
\zeta &=& \int d^{3}k dm \:\hat{\zeta}(m,k;\tau)e^{i\vec{k}.\vec{x}}
\eear
and recalling (\ref{solbulk}), for modes with $m \ne 0$ we have
\bear
\partial_{0}\hat{h} &=& -i\sqrt{m^2 + k^2}\hat{h} \quad , \\
\partial_{i} h_{\alpha\beta}|_{\bar{b}} &=& \int d^{3}k dm \:ik_{i}
\hat{h}(m,k;\tau)\: \epsilon_{\alpha\beta}(m,k)e^{i\vec{k}.\vec{x}} \quad ,\\ 
\partial_{4}\hat{h} &=& \hat{h}m\Bigl(\frac{2}{x} + \frac{Z_2'(x)}{Z_2(x)}
\Bigr)|_{x=ml/a} \quad . 
\eear
Substituting back in (\ref{Fij}) we find
\bear
\hat{F}^{i}_{j} &=& - \frac{\kappa}{2}\hat{\Sigma}^{i}_{j} -
\frac{1}{a^2}k^{j}k^{i}\hat{\zeta} \nonumber\\
&-& \frac{Hl}{2a}\Bigl[ik^{i}\epsilon_{0j} + ik^{j}\epsilon_{i0}  +
i\sqrt{m^2 + k^2}\epsilon_{ij} \Bigr]\hat{h} \nonumber \\ &-&
\frac{\sqrt{1+H^2l^2}}{2a} \epsilon_{ji}m\hat{h}\Bigl(\frac{2}{x} +
\frac{Z_2'(x)}{Z_2(x)} \Bigr)|_{x=ml/a}\quad .
\eear

Decomposing the above expression, using (\ref{stress}), the traceless-free
equation implies that 
\bear
\frac{\kappa}{2}\hat{\Sigma}^{1}_{2} &=&  -
\frac{\hat{h}}{2a}\epsilon_{12}\Bigl[iHl\sqrt{m^2 + k^2} +
m\sqrt{1+H^2l^2} \frac{Z_1(mla^{-1}(\tau))}{Z_2(mla^{-1}(\tau))}\Bigr] \quad ,
\nonumber \\ 
\frac{\kappa}{2}\hat{\Sigma}^{1}_{3} &=&  -
\frac{\hat{h}}{2a}\epsilon_{13}\Bigl[\frac{iHlm^{2}}{\sqrt{k^2+m^2}}+
m\sqrt{1+H^2l^2}\frac{Z_1(mla^{-1}(\tau))}{Z_2(mla^{-1}(\tau))}\Bigr] \quad ,
\nonumber\\ 
\frac{\kappa}{2}\hat{\Sigma}^{2}_{3} &=&  -
\frac{\hat{h}}{2a}\epsilon_{23}\Bigl[\frac{iHl m^2}{\sqrt{k^2+m^2}}   +
m\sqrt{1+H^2l^2}\frac{Z_1(mla^{-1}(\tau))}{Z_2(mla^{-1}(\tau))}\Bigr] \quad ,
\nonumber   
\eear
and
\bear
 \frac{1}{3}F &=& -\frac{\kappa}{2}\hat{\Sigma}^{1}_{1}  -
\frac{\hat{h}}{2a}\epsilon_{11}\Bigl[iHl\sqrt{m^2 + k^2} +
m\sqrt{1+H^2l^2} \frac{Z_1(mla^{-1}(\tau))}{Z_2(mla^{-1}(\tau))}\Bigr] \quad ,
\nonumber \\ 
  &=& -\frac{\kappa}{2}\hat{\Sigma}^{2}_{2}   -
\frac{\hat{h}}{2a}\epsilon_{22}\Bigl[iHl\sqrt{m^2 + k^2} +
m\sqrt{1+H^2l^2} \frac{Z_1(mla^{-1}(\tau))}{Z_2(mla^{-1}(\tau))}\Bigr] \quad ,
\nonumber\\  
  &=& -\frac{\kappa}{2}\hat{\Sigma}^{3}_{3}  -
\frac{\hat{h}}{2a}\epsilon_{33}\Bigl[\frac{iHl(m^2-k^2)}{\sqrt{k^2+m^2}} +
m\sqrt{1+H^2l^2} \frac{Z_1(mla^{-1}(\tau))}{Z_2(mla^{-1}(\tau))}\Bigr] -
\frac{k^2}{a^2}\hat{\zeta} \quad .\nonumber  
\eear

In dynamical universes without
the presence of an anisotropic stress, we claim that
\begin{itemize}
\item{ Massive modes are such that  $\epsilon_{11}(k,m)=
\epsilon_{22}(k,m)$. That means, from the transverse traceless-free condition} 
\bear
\epsilon_{11}(k,m) &=& \epsilon_{22}(k,m) =
-\frac{1}{2}\frac{m^2}{k^2+m^2}\epsilon_{33}, \\ 
\epsilon_{00}(k,m) &=& \frac{k^2}{k^2+m^2}\epsilon_{33}(k,m).
\eear

Thence, we find, recalling the form of $\hat{h}$,
\bear
\hat{\zeta}(m,k;\tau)
 &=& \frac{m^2l^2}{2k^2a(\tau)}\epsilon_{33}e^{-i\sqrt{k^2+m^2}T(\tau)}
\Bigl[\frac{iHl}{\sqrt{m^2+k^2}}\Bigl( k^{2} -
\frac{3m^2}{2}\Bigr)Z_{2}(mla^{-1}(\tau)) \nonumber \\ &-&
\frac{m}{(m^2+k^2)}\Bigl(\frac{3m^2}{2}+ k^2\Bigr)\sqrt{1+H^2l^2}
Z_1(mla^{-1}(\tau))\Bigr]\quad .
\eear

\item{ Since the Bessel's functions are linearly independent, we can not
have massive KK modes produced in the polarizations $\epsilon_{12}$,
$\epsilon_{23}$ and $\epsilon_{13}$. That means, for $m\ne 0$, } 
\bear
\epsilon_{12}(k,m) = \epsilon_{13}(k,m) = \epsilon_{23}(k,m) =
\epsilon_{01}(k,m) = \epsilon_{02}(k,m) = 0\quad ,
\eear
remaining just one degree of freedom, $\epsilon_{33}(k,m)$.
\end{itemize}

\section{Massless Mode}
For modes with $m = 0$ we  write
\bear
h_{\alpha\beta}|_{\bar{b}} &=& \int d^{3}k  \:\hat{h}(0,k;\tau)\:
\epsilon_{\alpha\beta}(0,k) e^{i\vec{k}.\vec{x}}\quad , \\
\zeta &=& \int d^{3}k \:\hat{\zeta}(0,k;\tau)e^{i\vec{k}.\vec{x}}\quad ,
\eear
leading to the result
\bear
\hat{F}^{i}_{j} &=& - \frac{\kappa}{2}\hat{\Sigma}^{i}_{j} -
\frac{1}{a^2}k^{j}k^{i}\hat{\zeta} -
\frac{Hl}{2a}\Bigl[ik^{i}\epsilon_{0j} + ik^{j}\epsilon_{i0}  +
ik\epsilon_{ij} \Bigr]\hat{h} \nonumber \\ &-& \frac{\sqrt{1+H^2l^2}}{2a}
\epsilon_{ji}\frac{4b_{0}l^{4}a(\tau)}{c_{0}a^{4}(\tau) +
b_{0}l^{4}}\hat{h}\quad .
\eear

Then, with $k^{i} = \delta^{i}_{3}k$, and the traceless-free equation we have
\bear
 \frac{\kappa}{2}\hat{\Sigma}^{1}_{2}  &=& -
\frac{\hat{h}}{2a}\epsilon_{12}\Bigl(iHlk+
\sqrt{1+H^2l^2}\frac{4b_{0}l^{4}a(\tau)}{c_{0}a^{4}(\tau) +
b_{0}l^{4}}\Bigr) \quad , \nonumber \\ 
 \frac{\kappa}{2}\hat{\Sigma}^{1}_{3} &=&  -
\frac{\hat{h}}{2a}\epsilon_{13}\Bigl(
\sqrt{1+H^2l^2}\frac{4b_{0}l^{4}a(\tau)}{c_{0}a^{4}(\tau) +
b_{0}l^{4}}\Bigr) \quad  , \nonumber \\ 
 \frac{\kappa}{2}\hat{\Sigma}^{2}_{3} &=& -
\frac{\hat{h}}{2a}\epsilon_{23}\Bigl(
\sqrt{1+H^2l^2}\frac{4b_{0}l^{4}a(\tau)}{c_{0}a^{4}(\tau) +
b_{0}l^{4}}\Bigr) \quad , \nonumber 
\eear
and
\bear
 \frac{1}{3}F  &=& -\frac{\kappa}{2}\hat{\Sigma}^{1}_{1} -
\frac{\hat{h}}{2a}\epsilon_{11}\Bigl(iHlk+
\sqrt{1+H^2l^2}\frac{4b_{0}l^{4}a(\tau)}{c_{0}a^{4}(\tau) +
b_{0}l^{4}}\Bigr) \quad , \nonumber \\ 
 &=& -\frac{\kappa}{2}\hat{\Sigma}^{2}_{2}  -
\frac{\hat{h}}{2a}\epsilon_{22}\Bigl( iHlk+
\sqrt{1+H^2l^2}\frac{4b_{0}l^{4}a(\tau)}{c_{0}a^{4}(\tau) +
b_{0}l^{4}}\Bigr) \quad , \nonumber \\ 
 &=& -\frac{\kappa}{2}\hat{\Sigma}^{3}_{3} -
\frac{\hat{h}}{2a}\epsilon_{33}\Bigl( -iHlk+
\sqrt{1+H^2l^2}\frac{4b_{0}l^{4}a(\tau)}{c_{0}a^{4}(\tau) +
b_{0}l^{4}}\Bigr) - \frac{k^{2}}{a^2}\hat{\zeta}\quad . \nonumber
\eear

From those equations, in the absence of an anisotropic stress we may claim that
\begin{itemize}
\item{From the transverse traceless-free condition, massless modes
are such that $\epsilon_{11} = - \epsilon_{22}$, thence
$\epsilon_{11} = \epsilon_{22} = 0$, $\hat{F} = 0$ and }
\bear  \hat{\zeta}
= - \frac{\hat{h}a}{2k^2}\epsilon_{33}\Bigl( -iHlk+
\sqrt{1+H^2l^2}\frac{4b_{0}l^{4}a(\tau)}{c_{0}a^{4}(\tau) +
b_{0}l^{4}}\Bigr)\quad .
\eear

\item{Massless modes are such that $\epsilon_{12}=0$ and we
must choose either $\epsilon_{23} =
\epsilon_{13} = 0$ or $b_{0} = 0$. The last choice corresponds to the
bounded zero mode advocated before. In this case, there still remain the
degrees of freedom $e_{33}$, $e_{23}$ and $e_{13}$ with } 
\bear  \hat{\zeta}
= \frac{iHl\hat{h}a}{2k}\epsilon_{33}\quad .
\eear

\end{itemize}

We can gather all results  under the following statements:

\begin{enumerate}

\item{There are, at most, three independent degrees of freedom of
polarization, denoted by $\epsilon_{33}(k,m)$, $\epsilon_{13}(k,m)$ and
$\epsilon_{23}(k,m)$, for $k^{i} = k\delta^{i}_{3}$. All the components
are written as} 
\bear
\epsilon_{11}(k,m) &=& \epsilon_{22}(k,m) =
-\frac{1}{2}\frac{m^2}{m^2+k^2}\epsilon_{33}(k,m) \quad ,\nonumber \\ 
\epsilon_{00}(k,m) &=& \frac{k^2}{m^2+k^2}\epsilon_{33}(k,m),\nonumber \\
\epsilon_{01}(k,m) &=& -\epsilon_{31}(k,0)\delta(m) \quad , \nonumber \\
\epsilon_{02}(k,m) &=& -\epsilon_{32}(k,0)\delta(m) \quad , \nonumber \\
\epsilon_{03}(k,m) &=& -\frac{k}{\sqrt{k^2+m^2}}\epsilon_{33}(k,m), \nonumber\\
\epsilon_{31}(k,m) &=& \epsilon_{31}(k,0) \delta(m) \quad , \quad
\epsilon_{32}(k,m) = \epsilon_{32}(k,0) \delta(m) \quad , \nonumber \\
\epsilon_{33}(k,m) &=& \epsilon_{33}(k,m) \quad , \quad 
\epsilon_{12}(k,m) = 0 \quad . \nonumber
\eear

\item{The perturbations $h_{\alpha\beta}$ can be written  as}
\bear
h_{\alpha\beta}(t,X^{4}) = \int d^{3}k dm
(mX^{4})^{2}\epsilon_{\alpha\beta}(k,m)N_{2}(mX^{4})e^{-i\sqrt{k^2+m^2}t +
ikx^{3}}\quad .
\eear
On the brane, it reads
\bear
h_{\alpha\beta}(\tau) = \int d^{3}k dm \: \frac{m^2l^2}{a^{2}(\tau)}
\epsilon_{\alpha\beta}(k,m)N_{2}(mla^{-1}(\tau))e^{-i\sqrt{k^2+m^2}T(\tau)
+ ikx^{3}} \quad . \nonumber \\ 
\eear

\item{Finally, the perturbation on the position of the brane is}
\bear
\zeta(\tau)
 &=& \int d^{3}k dm \:
\frac{m^2l^2}{2a(\tau)}\epsilon_{33}(k,m)e^{-i\sqrt{k^2+m^2}T(\tau) +
ikx^{3}} \times \nonumber \\
&\times&\Bigl[\frac{iHl}{\sqrt{m^2+k^2}}\Bigl( 1 -
\frac{3m^2}{2k^2}\Bigr)N_{2}(mla^{-1}(\tau)) - \nonumber \\ &-&
\frac{m}{(m^2+k^2)}\Bigl( 1+ \frac{3m^2}{2k^2}\Bigr)\sqrt{1+H^2l^2}
N_1(mla^{-1}(\tau))\Bigr]\quad .
\eear

\end{enumerate}



\acknowledgements{This work was partially supported by NNSF of China,
Ministry of Education of China and Ministry of Science and Technology of
China under Grant NKBRSFG19990754, 2003cb716300 as well as by FAPESP and
CNPQ, Brazil. The authors would also like to thank N. Deruelle to
discussions.} 


\end{document}